\documentstyle[12pt,epsfig,axodraw]{article}

\newcommand{\bea}{\begin{eqnarray}}
\newcommand{\eea}{\end{eqnarray}}

\newcommand{\bcen}{\begin{center}}
\newcommand{\ecen}{\end{center}}

\newcommand{\lsim}{\raisebox{-0.13cm}{~\shortstack{$<$ \\[-0.07cm] $\sim$}}~}
\newcommand{\gsim}{\raisebox{-0.13cm}{~\shortstack{$>$ \\[-0.07cm] $\sim$}}~}

\newcommand{\ra}{\rightarrow}

\newcommand{\s}{\\ \vspace*{-3mm} }
\newcommand{\nn}{\noindent}

\newcommand{\beq}{\begin{eqnarray}}
\newcommand{\eeq}{\end{eqnarray}}
\newcommand{\miss}{\not\hspace*{-1.8mm}E}

\begin{document}

\vspace*{1cm} 

\begin{flushright}
PM/01--06\\
May 2001\\
\end{flushright}

\vspace*{0.9cm}

\begin{center}

{\large\sc {\bf Scalar top quarks at the Run II of the Tevatron}}

\vspace*{0.4cm}

{\large\sc {\bf in the high $\tan \beta$ regime}}

\vspace{0.7cm}

{\sc A. Djouadi$^{1}$, M. Guchait$^{1}$ and Y. Mambrini$^{1,2}$ } 

\vspace{0.7cm}

$^1$ Laboratoire de Physique Math\'ematique et Th\'eorique, UMR5825--CNRS,\\
Universit\'e de Montpellier II, F--34095 Montpellier Cedex 5, France. 

\vspace*{3mm}

$^2$ CEA/DIF/DPTA/SPN, B.P. 12, F--91680 Bruy\`eres le Ch\^atel, France. 
\end{center}

\vspace*{1cm} 

\begin{abstract}
\nn We discuss the decays of the lightest scalar top quark $\tilde{t}_1$ in the
Minimal Supersymmetric extension of the Standard Model and show that the final
state $b\tau \nu_\tau \chi_1^0$, where $\chi_1^0$ is the lightest
supersymmetric particle, can be dominant in a significant area of the parameter
space, in particular in the high $\tan \beta$ regime. We then analyze the
prospects for discovering relatively light scalar top quarks accessible at the
Tevatron Run II in the $b\tau$ and missing energy channel.  
\end{abstract}

\newpage 

\subsection*{1. Introduction}

In the Minimal Supersymmetric (SUSY) extension of the Standard Model (MSSM)
\cite{MSSM}, the phenomenology of the spin-zero partners of the top quark is
rather special \cite{theses}. Because of the large $t$--quark Yukawa
coupling, the evolution from the high (unification) scale to the electroweak
scale of the soft--supersymmetry breaking scalar masses of left-- and
right--handed top squarks, $\tilde{t}_L$ and $\tilde{t}_R$, is different from
the one of the partners of the light fermions \cite{msugra}. In addition, the
two current eigenstates $\tilde{t}_L$ and $\tilde{t}_R$ mix very strongly, the
mixing being proportional to the fermion mass, leading to a possibly large mass
splitting between the two physical eigenstates $\tilde{t}_1$ and $\tilde{t}_2$
\cite{stop}.  The top squark $\tilde{t}_1$ can be therefore much lighter than
the other squarks and possibly lighter than the  top quark itself.  

If $\tilde{t}_1$ is lighter than the $t$--quark and the other SUSY particles
[in particular the lightest chargino $\chi_1^+$ and the charged and neutral
scalar leptons $\tilde{\ell}$ and $\tilde{\nu}$] and assuming R--parity
conservation with the lightest SUSY particle (LSP) being the neutralino
$\chi_1^0$, it will have only two decay modes. The first channel, which has
been used to search for $\tilde{t}_1$ at LEP \cite{stLEP} and at the Tevatron
\cite{stTEV} in the past, is the loop induced and flavor changing decay into a
charm quark and the LSP neutralino, $\tilde{t}_1 \to c \chi_1^0$ \cite{loop}. 
The second channel is the four--body decay mode into a $b$--quark, the LSP and
two massless fermions, $\tilde{t}_1 \to b \chi_1^0 f\bar{f}'$, which is
mediated by heavier SUSY particle exchange \cite{4body}.  The two modes are
of the same order of perturbation theory, i.e. ${\cal O} (\alpha^3)$, and thus
compete with each other.

However, $\tilde{t}_1$ might not be the next--to--lightest SUSY particle.  In
minimal Supergravity (mSUGRA) type models \cite{msugra}, where one assumes a
universal mass $m_0$ for the scalar fermions and a common mass $m_{1/2}$ for
the gauginos at the GUT scale, the scalar partner of the tau lepton,
$\tilde{\tau}_1$, can become rather light for large enough values of $\tan
\beta$ [the ratio of the vacuum expectation values of the two Higgs doublets
needed to break the electroweak symmetry in the MSSM] and $\mu$ [the
Higgs--higgsino mass parameter] and this might dramatically affect the pattern
of $\tilde{t}_1$ decays.

Indeed if $\tilde{\tau}_1$ is lighter than the scalar top quark, the
three--body channel $\tilde{t}_1 \to b \tilde{\tau}_1 \nu_\tau$ will be
kinematically accessible and would dominate the $c\chi_1^0$ decay mode
\cite{3body}.  For even larger $\tilde{t}_1$ masses, the two--body decay
channel $\tilde{t}_1 \to b \chi_1^+$ would open up and overwhelm all other
decays, and if $m_{\chi_1^+} \geq m_{\tilde{\tau}_1}$, the main decay mode of
the lightest chargino would be $\chi_1^+ \to \tilde{\tau}_1 \nu_\tau$ with
$\tilde{\tau}_1$ decaying into $\tau \chi_1^0$ final states \cite{2body}. In
fact, even if $m_{\tilde{\tau}_1} \gsim m_{\tilde{t}_1}$, the contribution of
the diagram with $\tilde{\tau}_1$ exchange in the four--body decay mode will be
large, since the virtuality of $\tilde{\tau}_1$ is smaller; the final state
$\tilde{t}_1 \to b \chi_1^0 \tau \nu_\tau$ would be then dominant.  

Thus, for large $\tan \beta$ values and for stop masses of the order of 100--200
GeV which are accessible at the Tevatron Run II, the dominant decay mode of 
$\tilde{t}_1$ could be into a $b$--quark, the LSP neutralino and  
$\tau \nu_\tau$ pairs, i.e. with a final state consisting of a $b$--quark, a 
$\tau$ lepton and the missing energy due to the undetected LSP and neutrino. 
This topology is quite different from the ones which have been used to search 
for top squark pairs at the Tevatron up to now, i.e. two acoplanar jets plus 
missing energy for the mode $\tilde{t}_1 \to c \chi_1^0$ \cite{stTEV}, one
lepton plus missing energy for the mode $\tilde{t}_1 \to b \chi_1^+ \to b 
\chi_1^0 W \to  b \chi_1^0 \nu + e/\mu$ [the other $W$ boson 
decaying hadronically], and two leptons, a $b\bar{b}$ pair and missing energy 
for the decay mode $\tilde{t}_1 \to b \ell \tilde{\nu}$ with $\ell =e/\mu$ 
\cite{stlepton}, where the sneutrino decays invisibly [the chargino is assumed
to be heavier] into the LSP neutralino and a neutrino.

In this note we will discuss the prospects for discovering the lightest top 
squark $\tilde{t}_1$ at the Tevatron Run II in the decay channel $b\tau +\miss$.
After summarizing the decay branching ratios of $\tilde{t}_1$, we will discuss
the cross sections of the signal $p\bar{p} \to \tilde{t}_1 \tilde{t}_1^* \to b
\bar{b} \tau^+ \tau^- + \miss$, compared to those of the main background
originating from top--quark pair production which has the same topology,
$p\bar{p} \to t \bar{t} \to bW^+ \bar{b}W^- \to b \bar{b} \tau^+ \tau^- \nu
\bar{\nu}$. We will show that by tagging one $b$ quark and by requiring one
$\tau$ lepton to decay leptonically and the other hadronically, and after
suitable selections cuts, the discovery of top squarks with masses between 100
and 200 GeV is possible at the Tevatron, with a center of mass energy of 2 TeV 
and an integrated luminosity $\int {\cal L} \sim 20$ fb$^{-1}$.  

\subsection*{2. Scalar top decays branching ratios}

If top squarks are heavy enough, their main decay modes will be into top
quarks and neutralinos, $\tilde{t}_i \ra t\chi^0_j$ [$j$=1--4], and/or bottom
quarks and charginos, $\tilde{t}_i \ra b\chi^+_j$ [$j$=1--2].  If these modes
are kinematically not accessible, the lightest top squark can decay into a
charm quark and the lightest neutralino, $\tilde{t}_1  \ra c \chi_1^0$
\cite{loop}. This mode is mediated by one--loop diagrams: vertex diagrams as
well as squark and quark self--energy diagrams, where bottom squarks,
charginos, charged $W$ and Higgs bosons are running in the loops. The flavor
transition $b \ra c$ occurs through the charged currents. Adding the various
contributions, a divergence is left out which must be subtracted by adding a
counterterm to the scalar self--mass diagrams. When working in an mSUGRA
framework where the squark masses are unified at the GUT scale, the divergence
is subtracted using a soft--counterterm at $\Lambda_{\rm GUT}$, generating a
large residual logarithm $\log(\Lambda^2_{\rm GUT}/M_W^2) \sim 65$ in the
amplitude. This logarithm gives the leading contribution to the $\tilde{t}_1
\ra c \chi_1^0$ amplitude and makes the decay width rather large. [The decay
width is though suppressed by the CKM matrix element $V_{cb} \sim 0.05$ and the
(running) $b$ quark mass squared $m_b^2 \sim (3$ GeV$)^2$].  

However, there are scenarii in which the decay rate $\Gamma(\tilde{t}_1 \ra c
\chi_1^0$) can be rather small. First, the large logarithm $\log
\left(\Lambda_{\rm GUT}^2/ M_W^2 \right) \sim 65$ appears only because the
choice of the renormalization condition is made at $\Lambda_{\rm GUT}$, but in
a general MSSM where the squark masses are not unified at some very high scale,
one could chose a low energy counterterm; in this case no large logarithm would
appear.  In addition, if the lightest top squark is a pure right--handed state
[as favored by the constrains \cite{PDG} from high--precision electroweak
data], the amplitude involves only one component which can be made small by
choosing tiny values of the trilinear coupling $A_b$ and/or large values of the
(common) SUSY--breaking scalar mass $\tilde{m}_q$. Finally, even in the
presence of stop mixing and for a given choice of MSSM parameters, large
cancellations can occur between the various terms in the loop amplitude.
Thus, the decay rate $\Gamma (\tilde{t}_1 \ra c \chi_1^0)$ might be very small,
opening the possibility for the three-- or four--body decay modes to dominate. 

The four--body decay mode $\tilde{t}_1 \ra b \chi_1^0 f \bar{f}'$, proceeds 
through several diagrams, as shown in Fig.~1. There are first the $W$--boson 
exchange diagrams with virtual $\tilde{t}, \tilde{b}$ and $\chi^\pm_{1,2}$ 
states, a similar set of diagrams is obtained by replacing the $W$--boson 
by the $H^+$ boson and a third type of diagrams consists of up-- and down--type 
slepton and first/second generation squark exchanges. The decay rate has
been calculated in Ref.~\cite{4body} taking into account all diagrams and 
interferences. The various contributions can be summarized as follows.  

$i)$ Because in the MSSM, $H^\pm$ has a mass larger than $M_W$ and
has tiny Yukawa couplings to light fermions, it does not give rise to large 
contributions. The squark exchange diagrams give also small contributions 
since squarks are expected to be much heavier than the $\tilde{t}_1$ state, 
$m_{\tilde{q}} \gsim {\cal O}(250)$ GeV \cite{PDG}. The contribution of the 
diagram with an exchanged $t$--quark is only important if $m_{\tilde{t}_1}$ is 
of the order of $m_t+m_{\chi_1^0} \gsim {\cal O} (250$ GeV). 
$ii)$ A significant contribution to the four--body decay mode would come from
the first diagram in Fig.~1, when the virtuality of the chargino is not too
large. In particular, for an exchanged $\chi_1^+$ with a mass not much larger
than the present experimental bound, $m_{\chi_1^+} \gsim 100$ GeV \cite{PDG}, 
the decay width can be substantial even for top squark masses of the order of 
100 GeV.  
$iii)$ In contrast to the exchange of squarks, slepton exchange diagrams might 
give substantial contributions, since masses $m_{\tilde{\ell}} \sim {\cal O}
$(100 GeV) are still experimentally allowed \cite{PDG}. In fact, when the 
difference 
between $m_{\tilde{t}_1}$ and $m_{\chi_1^+}, m_{\tilde{l}}$ is not large, the 
third diagram in Fig.~1 will give the dominant contribution to the four--body 
decay mode.  In particular, if the $\tau$ slepton is rather light, the dominant
final state will be $\tilde{t}_1 \ra b \chi_1^0 \tau \nu$. 

\begin{picture}(1000,170)(0,0)
\Text(200,10)[]{\it Figure 1: Generic Feynman diagrams contributing to the 
decay $ \tilde{t}_1 \ra b\chi_1^0 f\bar{f}'$.}
\Text(0,120)[]{$\tilde{t}_1$}
\DashArrowLine(10,120)(40,120){4}{}
\ArrowLine(70,150)(40,120)
\Text(77,150)[]{$b$}
\ArrowLine(40,120)(60,90)
\Text(35,100)[]{$\chi_i^+$}
\ArrowLine(60,90)(90,120)
\Text(100,120)[]{$\chi_0$}
\Photon(80,65)(60,90){4}{8}
\Text(55,75)[]{$W$}
\ArrowLine(80,65)(110,95)
\Text(115,90)[]{$f$}
\ArrowLine(110,35)(80,65)
\Text(115,35)[]{$\overline{f}'$}
\Text(150,120)[]{$\tilde{t}_1$}
\DashArrowLine(160,120)(190,120){4}{}
\ArrowLine(220,150)(190,120)
\Text(227,150)[]{$b$}
\ArrowLine(190,120)(210,90)
\Text(185,100)[]{$\chi_i^+$}
\ArrowLine(210,90)(240,120)
\Text(250,120)[]{$\chi_0$}
\DashArrowLine(230,65)(210,90){4}{}
\Text(205,75)[]{$H^\pm$}
\ArrowLine(230,65)(260,95)
\Text(265,90)[]{$f$}
\ArrowLine(260,35)(230,65)
\Text(265,35)[]{$\overline{f}'$}
\Text(300,120)[]{$\tilde{t}_1$}
\DashArrowLine(310,120)(340,120){4}{}
\ArrowLine(340,120)(370,150)
\Text(377,150)[]{$b$}
\ArrowLine(340,120)(360,90)
\Text(335,100)[]{$\chi^+_i$}
\ArrowLine(390,120)(360,90)
\Text(400,120)[]{$f$}
\DashArrowLine(380,65)(360,90){4}{}
\Text(360,75)[]{$\tilde{f}^{*,}$}
\ArrowLine(380,65)(410,95)
\Text(415,90)[]{$\bar{f}'$}
\ArrowLine(410,35)(380,65)
\Text(418,35)[]{$\chi_0$}
\end{picture}


For larger stop masses, when the $W$ boson, the $H^\pm$ boson and/or the
sfermion $\tilde{f}^*$ in Fig.~1 are kinematically allowed to be on
mass--shell, one will have a three--body decay mode of the $\tilde{t}_1$. 
These modes have been discussed earlier in Refs.~\cite{3body} but the case of
the $\tilde{\tau}_1$ state, which might be rather light as mentioned
previously, has not been discussed explicitly. However, this is potentially
the dominant mode since the other modes can be strongly  suppressed: the
$H^\pm$ boson is expected to be rather heavy and its couplings to fermions
[except from $tb$ states which are not kinematically accessible] are rather
tiny, and if the lightest chargino and neutralino are gaugino like [as is
usually the case in mSUGRA--type models] the $W \chi_1^0 \chi_1^+$ coupling is
very small. Thus, even if the three channels are present at the same time, the
decay $\tilde{t}_1 \to b \tilde{\tau}_1 \nu$ with the subsequent decay
$\tilde{\tau}_1 \to \tau \chi_1^0$ [which is the only possible channel since
$m_{\chi_1^+} > m_{\tilde{\tau}_1}$ and the decay $\tilde{\tau}_1 \to \nu_\tau
\chi_1^+$ is kinematically shut] could be the dominant decay mode of the 
lightest top squark.  

Finally, if $m_{\tilde{t}_1} > m_{\chi_1^+} + m_b$, the two--body decay channel
$\tilde{t}_1 \to b\chi_1^+$ will be kinematically accessible and would dominate
all other possible modes. But if $\tilde{\tau}_1$ is light, the chargino will
dominantly decay into $\tilde{\tau}_1 \nu_\tau$ pairs \cite{2body}.  The other
possible decay mode $\chi_1^+ \to \chi_1^0 W$, if accessible, would be
suppressed by the small $W \chi_1^0 \chi_1^+$ coupling which vanishes for pure
gaugino--like charginos and neutralinos. For even higher masses,
$m_{\tilde{t}_1} \gsim {\cal O} (250)$ GeV, the decay channel $\tilde{t}_1 \to
t \chi_1^0$ opens up, and would compete with the previous decay mode.  

Thus in all these situations, the possibly dominant decay mode of the lightest
scalar top quark,  with a mass below $m_t + m_{\chi_1^0} \gsim 200$ GeV, would 
be into $b \chi_1^0 \tau \nu$ final states. This is illustrated in Fig.~2 where
we show the branching  ratio for this final state as a function of
$m_{\tilde{t}_1}$ for two values of $\tan \beta=5$ and 20 in an approximate
mSUGRA--type model. We have assumed gaugino mass unification and a common mass 
$m_0$ for all scalar fermion at the GUT scale, except for $m_{\tilde{t}_R
}$ which is varied to obtain $\tilde{t}_1$ masses between 100 and 200 GeV. The 
parameter $\mu$ is fixed to a high value, $\mu=750$ GeV, to generate a large 
mixing in the $\tilde{\tau}$ sector to obtain a light $\tilde{\tau}_1$ state; 
in this case the lightest chargino and neutralino are wino-- and bino--like with
masses, $m_{\chi_1^+} \simeq 2 m_{\chi_1^0} =M_2$. The trilinear couplings
$A_{t, b}$ are fixed to values of ${\cal O}(100$ GeV).  With the inputs
$M_2=170$ GeV, $m_0=170 \, (100)$ GeV for $\tan \beta=20\, (5)$, one obtains 
for the chargino and $\tau$ slepton masses: $m_{\chi_1^+} \simeq 2m_{\chi_1^0} 
\sim 165$ GeV and $m_{\tilde{\tau}_1}\sim 130$ GeV. All sparticles [as well as 
the $H^\pm$ boson] have large enough masses not to affect the decay branching
ratio, except for the charged sleptons $\tilde{e}, \tilde{\mu}$ and the
sneutrinos $\tilde{\nu}$  which have masses only slightly above $m_{\chi_1^+}$
in the $\tan \beta=5$ scenario.

As can be seen, for $\tan \beta=20$, the final state $b \chi_1^0 \tau \nu$ is
dominant in almost the entire range of $m_{\tilde{t}_1}$, in particular when
the three--body decay mode $\tilde{t}_1 \to b \tilde{\tau}_1 \nu$ and the
two--body decay mode $\tilde{t}_1 \to b \chi_1^+$ open up [the spikes
correspond to the opening of these channels, the finite widths of the virtual
particles not being included]. For the four body--decay mode, BR($\tilde {t}_1
\to b \chi_1^0 \tau \nu$) reaches values of the order of $50\%$ for
$m_{\tilde{t}_1} \gsim 100$ GeV; below this value, BR($\tilde{t}_1 \to c 
\chi_1^0$) [which is enhanced by ${\rm log}^2 (\Lambda^2_{\rm GUT} /M_W^2)$ 
and a factor $\tan^2 \beta$] is dominant because of the larger phase space 
[here $m_{\tilde{t}_1} \sim  m_b + m_{\chi_1^0}$].  For the small value $\tan
\beta=5$, the decay  $\tilde {t}_1 \to b \chi_1^0 \tau \nu$ is dominant only
for masses $m_{\tilde{t}_1} \gsim 150$ when the channel $\tilde{t}_1 \to b
\chi_1^+$ is about to open up. Below this value, there is a competition from
the channels with the $W$ exchange diagram [here, $m_{\chi_1^+}\sim m_{\chi_1^0}
+M_W$] and with the exchange of the other sleptons [which have masses close to
$m_{\chi_1^+}$]. In this case, BR($\tilde {t}_1 \to b \chi_1^0 \tau \nu$) is
too small, and one has to rely on the decays involving $e,\mu$ final states
\cite{stlepton,lykken}. 

\setcounter{figure}{1}
\begin{figure}[htbp]
\vspace*{-5.5cm}
\hspace*{-3.5cm}
\mbox{\psfig{figure=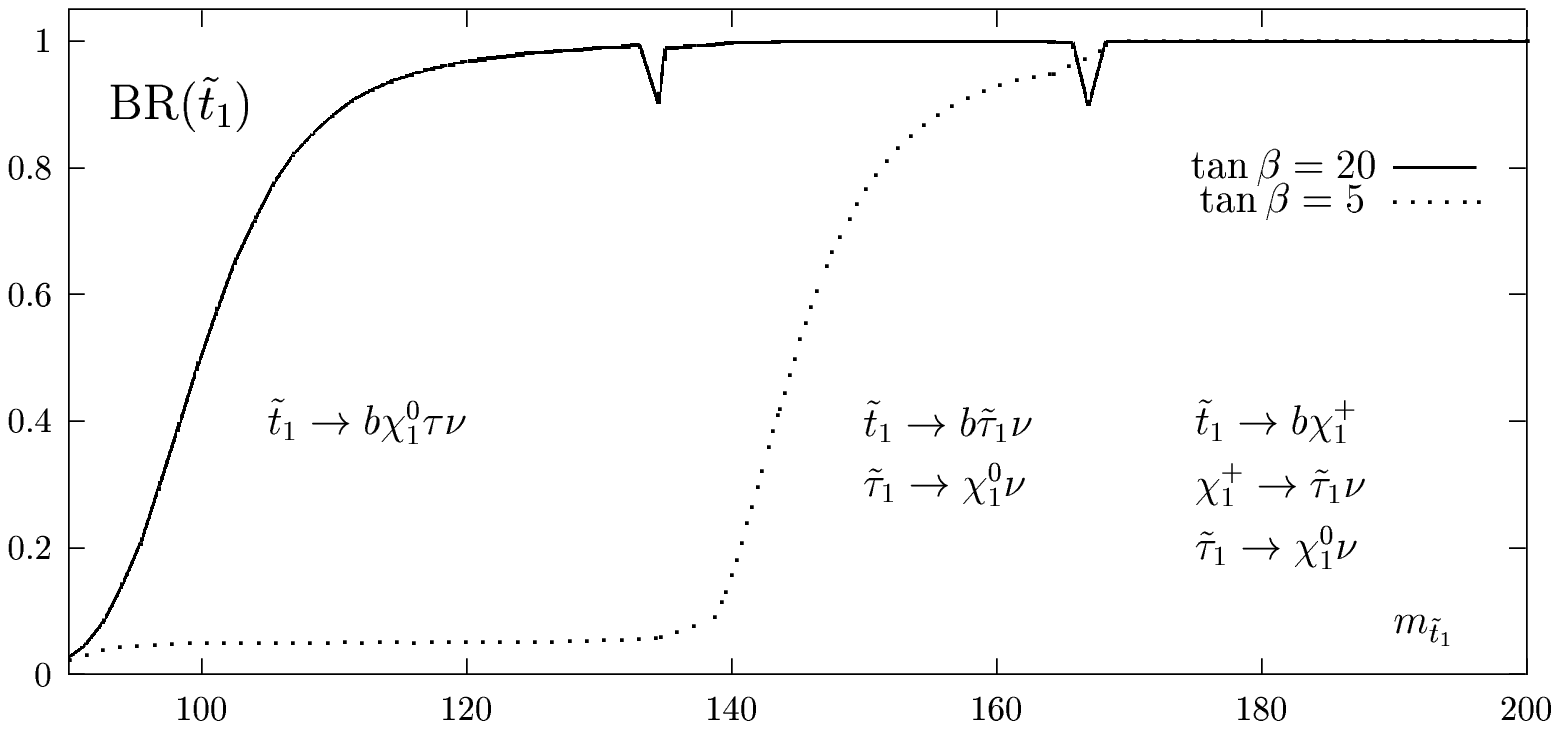,width=20cm}}
\vspace*{-16.7cm}
\caption[]{\it The branching ratio BR$(\tilde{t}_1 \to b \chi^0_1 \tau \nu)$
versus $m_{\tilde{t}_1}$ for $\tan \beta= 5$ and 20. The SUSY parameters are 
such that $m_{\chi_1^+} \simeq 2 m_{\chi_1^0} \simeq 165$ GeV and 
$m_{\tilde{\tau}_1} \simeq 130$ GeV.}
\end{figure}

\subsection*{3. Scalar top production: signal and backgrounds} 

Top squarks are produced in hadronic collisions via quark--antiquark
annihilation and gluon--gluon fusion \cite{prod1}. The total cross section at
the Tevatron with $\sqrt{s}=2$ TeV is of the order of $\sigma( p\bar{p} \to
\tilde{t}_1 \tilde{t}_1^*) \sim 15$ to 0.3 pb for $\tilde{t}_1$ masses in the
range of 100 to 200 GeV. A factor $K \sim 1.3$, to take into account the
next--to--leading order QCD corrections \cite{prod2}, has been included.  The
renormalization scale has been chosen at $\mu^2 = m_{\tilde{t}_1}^2$ and the
CTEQ3L \cite{CTEQ} parameterization of the parton densities has been used.  This
has to be compared with the production cross section of top quark pairs,
$\sigma(p\bar{p} \to t\bar{t}) \sim 8$ pb, which includes the factor $K
\sim 1.4$ for QCD corrections \cite{ttbar}.  

In the following, we will chose the scenario discussed in the previous section,
i.e. with the lightest chargino, neutralino and $\tilde{\tau}_1$  masses being
$m_{\chi_1^+} \simeq 2 m_{\chi_1^0} \simeq 165$ GeV and $m_{\tilde{\tau}_1}
\simeq 130$ GeV, and assume that top squarks will decay into $b \chi_1^0 \tau
\nu_\tau$ final states with a branching ratio close to unity. Therefore, pair
production of top squarks leads to final states containing two $b$--quarks, a
$\tau^+ \tau^-$ pair and missing transverse energy.

This final state can be triggered by selecting $b\tau_{\ell} \tau_h + \miss_T$, 
where we assume that one $\tau$ lepton decays leptonically, with a branching 
ratio BR$(\tau^\pm \to e^\pm+\mu^\pm) \simeq 35\%$, while the other decays 
hadronically with BR$(\tau^\pm \to {\rm hadrons}) \, \simeq 65\%$. The hadronic
$\tau$ jets can be tagged as narrow jets with their three main decay modes
being $\tau^\pm \rightarrow \pi^\pm \nu_\tau \, (18\%) , \, \tau^\pm
\rightarrow \rho^\pm \nu_\tau \, (24\%)$ and $\tau^\pm \rightarrow a_1^\pm
\nu_\tau \, (7.5\%)$.  Each vector meson ($a_1,\rho$) decays hadronically in
one charged pion and other neutral pions signaling 1--prong decay modes in the
detector.  The decay distribution is normalized according to the decay modes
following the prescription  given in Ref.~\cite{Hagiwara}.  The cross section
will be diluted by the branching ratio suppression for $\tau \tau 
\rightarrow \ell \tau_h$ by $B_{\ell \tau_h} = 0.35 \times 2 \times (2/3) 
\simeq 45\%$ for $\ell = e, \mu$. Of course, to increase the statistics, one
could also use the topology where both tau leptons decay hadronically which 
has a larger branching ratio than the $\tau_h \tau_{\ell}$ mode; this channel, 
although viable \cite{private}, will not be discussed in this note. 

We have calculated the signal [and main background] cross sections by using a 
simple parton--level Monte--Carlo simulation, i.e. without taking into account
fragmentation effects of jets.  For for the selection of the signal events, we
use the following set of cuts for the transverse momentum $p_T$, the rapidity
$\eta$ and the lepton/jet isolation cut defined as $\Delta R=\sqrt{(\Delta
\eta )^2 + (\Delta \phi)^2}$ with $\Delta \eta$ and $\Delta \phi$ being,
respectively, the difference of rapidity and azimuthal angle between the
leptons and any of the jets: \s

\nn 1. Lepton selection: $p_T^\ell >7$ GeV and $|\eta_\ell|< 2$ with $\Delta 
R >$0.4. 

\nn 2. $\tau$ jet selection: $p_T^\tau >$10 GeV and $|\eta_\tau|<4$ with 
$\Delta R>$~0.5. 

\nn 3. Missing energy selection: ${p\!\!\!/}_T >$~15 GeV. 

\nn 4. $b$--jet selection:  number of $b$-jets $\ge$~1 with $p_T^b>$~15 GeV, 
$|\eta_b|<$ 4 and $\Delta R >$0.7. \s

\begin{table}[!htb]
\begin{center}
\renewcommand{\arraystretch}{1.35}
\begin{tabular}{|c|c|c|c|c|c||c|}
\hline
Cut $\downarrow$ \, / $m_{\tilde t_1} \rightarrow $  & 100  & 120 & 150 
& 180 & 200 & \ \ \ $t\bar t$ \ \ \ \\ 
\hline 
0 & 7.2 & 2.7  & .76  & .25 & .13 &.26 \\ 
1 & .39 &.57  & .52 &.17  & .09 &.24 \\
2 & .20  &.09  & .36 & .12 &.06 & .14 \\
3 & .006 &.06  & .30 & .10 &.058 &.14 \\
4 & .004 & .044 & .16 & .07 & .057& .14 \\ \hline
$H_T <$ 100 GeV & .004 &.043 & .10 &.046& .032& .016\\
\hline
\end{tabular}
\vspace*{-2mm}
\end{center}
\caption[]{\it The $\tilde{t}_1 \tilde{t}_1^*$  signal and $t\bar{t}$ 
background event cross sections [in pb] after the selection cuts 0--4 and 
the $H_T$ cut have been applied.}
\end{table}

The cross section for the signal after successively applying these cuts are
shown in Table 1 for several values of $m_{\tilde{t}_1}$. The cut ``$0$" stands
for the total production cross section times the branching ratio for one $\tau$
decaying leptonically and the other decaying hadronically [BR $\sim 45\%$]. 
The cuts are much harder in the case of light top squarks than for heavier
ones: while the cross section is suppressed by three orders of magnitude for
$m_{\tilde{t}_1} \sim 100$ GeV, there is only a factor of two suppression for
$m_{\tilde{t}_1} \sim 200$ GeV. This is mainly due to the fact that, in the 
chosen scenario,  the mass of the LSP, $m_{\chi_1^0} \sim 
80$ GeV, is too close to $m_{\tilde{t}_1}$ so that the $b$ quarks and $\tau$ 
leptons are much softer in the former case. For lighter neutralinos or
heavier top squarks, the cuts are much less severe.

The main sources of background, besides $t\bar t$ pair production, are gauge
boson pair production: $p \bar{p} \to W^+W^-, W^\pm Z, ZZ$ and $Z \gamma$, plus
eventually some QCD jets. All these background processes can be removed by
requiring a fair amount of missing transverse energy as well as at least one
$b$--quark jet and two tau leptons, since the probabilities for misidentifying
jets with $b$--quarks and $\tau$--leptons is very small \cite{fake}. For
instance, in the case of $W$--pair production which has the largest cross
section, $\sigma(p \bar{p} \to W^+W^-) \sim 11$ pb, requiring a final state
$\ell \nu q\bar{q}'$ plus an additional soft QCD jet faking a $\tau$--jet, the
cross section drops to the level of 0.2 pb after imposing the selection cuts
discussed previously. Assuming probabilities of misidentifying the jets as
$\tau_h$ and $b$--quarks of a few percent \cite{fake}, the cross section will
be suppressed by at least three orders of magnitude, well below the level 
of the signal cross section even for $m_{\tilde{t}_1} \sim 100$ GeV.

Therefore, one should be left only with the $t\bar{t}$ background process. The
total production cross section, $\sigma (p \bar{p} \to t\bar{t}) \sim 8$ pb,
should be multiplied by the total branching suppression factor which leads to
the relevant $b\bar{b} \tau_h \ell +\miss$ final state [in $t\bar{t} \to
b\bar{b} W^+ W^-$ one $W$ boson decays into $e/\mu$ while the other decays into
a $\tau$ lepton which then decays hadronically], that is $B_{t \bar t} \simeq
(2/9) \times (1/9) \times (2/3) \times 2 \simeq  0.032$, leading to a cross
section of $\sigma \sim 0.26$ pb in this final state.  

The cross sections times branching ratio for the $t\bar t$ background, after
the successive cuts discussed above, are also shown in Table 1. As can be seen,
it is possibly larger than the signal cross section, especially for light
stops.  To suppress further this background we apply the cut: $H_T <$ 100 GeV
where $H_T = |p^T_\ell | + | p^T_{\tau_h} | + | {p\!\!\!/}_T |$.  Since all
these transverse momenta for $t \bar t$ production are harder than in the case
of the signal events for range of stop masses accessible at the Tevatron, this
reduces the $t \bar t$  background significantly.  As for example, after this
cut, the $t \bar t$ cross section is suppressed by one order of magnitude and
finally leads to a background cross section of 16 fb. In the case of the signal,
this cut is harmless as shown in Table 1.  

The final output is displayed in Fig.~3, where the signal cross
section\footnote{The spikes are again due to the opening of the three--body and
two--body decay channels, where the phase space for the decays $\tilde{t}_1 \to
b \tilde{\tau}_1 \nu$ and $\tilde{t}_1 \to b \chi_1^+$ is very small, leading
to very soft $b$--quark jets. This is quite artificial since, if the finite
decay widths of the chargino and $\tau$ slepton are included in the decay
amplitudes, there will be a smooth transition between the 4--body to the
3--body to the 2--body decay branching ratios. These decay widths have not been
included in the analysis since for the integration of the complicated 4--body
phase space, we use a Monte--Carlo which is not enough precise to resolve the
tiny total widths [compared to the masses] of the exchanged SUSY particles.}
after all cuts is shown as a function of $m_{\tilde{t}_1}$. Ones sees that for
$m_{\tilde{t}_1} \gsim 100$ GeV, $\sigma (\tilde{t}_1 \tilde{t}_1^* \to
b\bar{b} \tau_{\ell} \tau_h +\miss )$ is larger than 10 fb and reaches a
maximum of $\sim 100$ fb for $\tilde{t}_1$ masses between 120 and 170 GeV. The
$t\bar{t}$ background, $\sigma(t\bar{t})\sim 16$ fb, is much smaller after all
cuts have been applied.  One has then to multiply these numbers by the
$b$--quark and $\tau$ lepton tagging efficiencies, which are expected to be of
the order of 50\% \cite{fake,private} and which leads to a further
suppression\footnote{In fact, there is no need for $b$--tagging since with 
the cut 4, $p_T^{\rm jet} >15$ GeV, the backgrounds
with gauge boson production will be suppressed to a negligible level 
\cite{private}. However, since very efficient micro--vertex detectors will be
available in both CDF and D0, it is safer to require the presence of a 
$b$--jet.} of the signal events by a factor of 4. This means that more than 10
(100) events in this topology can be collected for stop masses in the range 120
GeV $\lsim m_{\tilde{t}_1} \lsim 180$ GeV, with an integrated luminosity of 2
(20) fb$^{-1}$ as expected in the two runs of the Tevatron\footnote{As
mentioned previously, the problem in the lower stop mass range is mainly due to
the small $m_{\tilde{t}_1} - m_{\chi_1^0}$ difference which leads to very soft
$b$ quarks and $\tau$ leptons that do not pass the selection cuts. For lighter
neutralinos, the phase space would be larger and the signal cross section can
be enhanced to a visible level. For the high stop mass range, $m_{\tilde{t}_1}
\gsim 200$ GeV, one is simply limited by the smallness of the production cross
section.}.  Therefore, one could hope to observe top squark events in this
channel at the Tevatron, at least in the high--luminosity option.

\begin{figure}[htbp]
\vspace*{-5.2cm}
\hspace*{-4.4cm}
\mbox{\psfig{figure=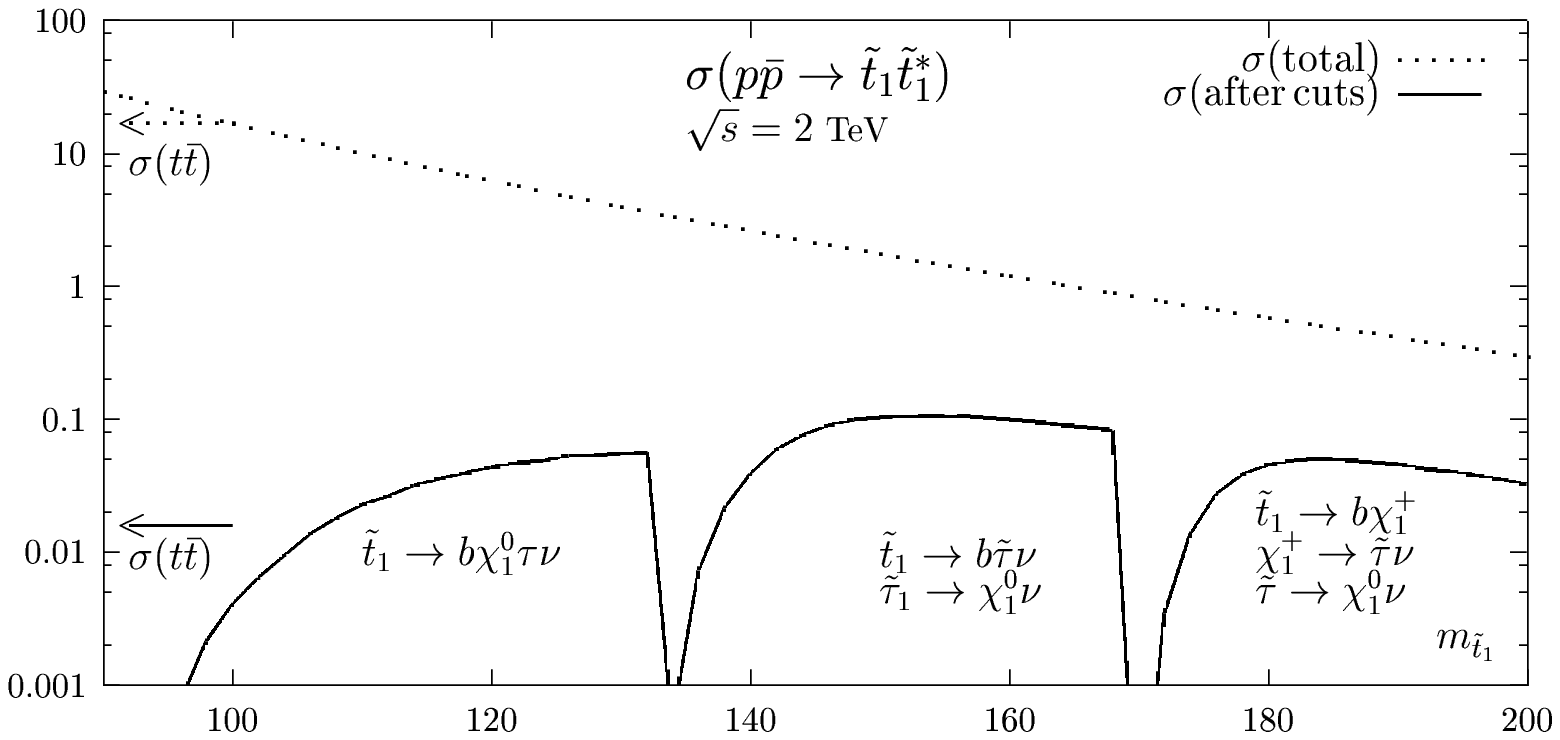,width=21cm}}
\vspace*{-17.5cm}
\caption[]{\it The cross sections for $p\bar{p} \to \tilde{t}_1 
\tilde{t}_1^*$ in pb at the Tevatron as a function of $m_{\tilde{t}_1}$:
the dashed lines are the total cross section and the solid lines for the
cross section after selection cuts (see text). The cross
sections for $t\bar{t}$ production are indicated by the arrows.}
\end{figure}

\subsection*{4. Discussions and conclusions} 

We have shown that in the large $\tan \beta$ regime, scalar top quarks that are
accessible at the Run II of the Tevatron [i.e. with masses in the range of 100 
to 200 GeV] might decay predominantly into bottom quarks, $\tau$ leptons and a 
large amount of missing energy due to the escaping neutrinos and neutralinos. 
This is due to the fact that $\tau$ sleptons are also relatively light: either
the decay channels $\tilde{t}_1 \to b \tilde{\tau}_1 \nu_\tau$ and $\tilde{t}_1
\to b \chi_1^+ \to b \tilde{\tau}_1 \nu_\tau$ are kinematically open, or the
virtuality of the exchanged $\tau$ sleptons in the four--body decay channel 
is small making the final state, $\tilde{t}_1 \to b \chi_1^0 \tau \nu$ 
dominant. 

We have performed a crude estimate of the prospects for discovering such a
light scalar top quark in the $p \bar{p} \to \tilde{t}_1 \tilde{t}_1^* \to
b\bar{b} \tau^+ \tau^- + \miss$ channel at the Run II of the Tevatron. 
Requiring one of the $\tau$ leptons to decay leptonically and the other
hadronically and the tagging of one $b$--quark [with reasonable efficiencies],
and applying rather loose cuts to select the signal, we have shown that this
signal can give a substantial number of events, in particular in the high
luminosity option $\int {\cal L} \sim 20$ fb$^{-1}$, and that it can stand over
the background, which is dominantly due to top quark pair production in the
channel $p \bar{p} \to t \bar{t} \to b\bar{b} \tau \ell + \nu \bar{\nu}$.  A
detailed analysis taking into account all backgrounds, selection and detection
efficiencies in a more realistic way, which is beyond the scope of this short
note, is nevertheless required to assess in which part of the MSSM parameter
space this final state can be isolated experimentally.  \bigskip

\nn {\bf Acknowledgments}: We thank  Gregorio Bernardi and Aurore Savoy--Navarro
for discussions.  M.G. thanks the CNRS for financial support and the LPMT for
the hospitality extended to him. Y.M. is supported by a MNERT fellowship and
thanks the members of the SPN for the facilities accorded to him.


\begin{thebibliography}{99}

\bibitem{MSSM} For reviews of the MSSM, see e.g: H. E. Haber and G. Kane, 
Phys. Rep. 117 (1985) 75; M. Drees and S. Martin, CLTP Report (1995) and 
hep-ph/9504324. 
%
\bibitem{theses} For reviews, see for instance: W. Porod, Doctoral thesis, 
(Vienna U.), hep-ph/9804208; T. Plehn, PhD. Thesis (Hamburg U.), 
hep-ph/9809319; S. Kraml, Doctoral thesis (Vienna, OAW), hep-ph/9903257.

%
\bibitem{msugra} For a review of mSUGRA and for the physics implications at 
the Tevatron Run II, see: S. Abel et al., Report of the ``SUGRA" working group 
for ``RUNII at the Tevatron", hep-ph/0003154.  
%
\bibitem{stop} J. Ellis and S. Rudaz, Phys. Lett. B128 (1983) 248; M. Drees 
and K. Hikasa, Phys. Lett. B252 (1990) 127; A. Bartl et al., hep--ph/9709252. 
%
\bibitem{stLEP} ALEPH Collaboration, Phys. Lett. B469 (1999) 303; DELPHI
Collaboration, Phys. Lett. B496 (2000) 59; L3 Collaboration, Phys. Lett. B471
(1999) 308; OPAL Collaboration, Phys. Lett. B456 (1999) 95; for a summary on
stop searches at LEP, see: S. Rosier--Lees et al., hep--ph/9901246.  

%
\bibitem{stTEV} 
D0 Collaboration (S. Abachi et al.), Phys. Rev. Lett. 76 (1996) 2222; CDF Collaboration (T. Affolder et al.), Phys. Rev. Lett. 84 (2000) 5704;
for a summary on stop searches at the Tevatron, see: A. Savoy-Navarro for the 
CDF and D0 Collaborations, Report FERMILAB-CONF-99-281-E (Nov. 1999). 
%
\bibitem{loop} K.I. Hikasa and M. Kobayashi, Phys. Rev. D36 (1987) 724. 
%
\bibitem{4body} C. Boehm, A. Djouadi and Y. Mambrini, Phys. Rev. D61 
(2000) 095006. 
%
\bibitem{3body} W. Porod and T. Wohrmann, Phys. Rev. D55 (1997) 2907; W. Porod,
Phys. Rev. D59 (1999) 095009;  A. Datta, M. Guchait and K.K. Jeong,  Int. J. 
Mod. Phys. A14 (1999) 2239; A. Djouadi and Y. Mambrini, 
 Phys. Rev. D63 (2001) 115005 (hep-ph/0011364).
%
\bibitem{2body} H. Baer, C.H. Chen, M. Drees, F. Paige and X. Tata, Phys. Rev.
D59 (1999) 055014 and Phys. Rev. Lett. 79 (1997) 986; A. Bartl et al., Phys.
Lett. B435 (1998) 118; A. Djouadi and Y. Mambrini, Phys. Lett. B493 (2000) 120;
A. Djouadi, Y. Mambrini and M. Muhlleitner, hep-ph/0104115 (EPJC, to appear). 
%
\bibitem{stlepton} CDF Collaboration (T. Affolder et al.), Phys. Rev. Lett. 84
(2000) 5273; V. Buescher (for the D0 collaboration), XXXVI Rencontres de
Moriond on Electroweak interactions and Unified Theories, 2001; B. Olivier, 
PhD thesis, Universit\'es Paris VI et VII (April 2001). 
%
\bibitem{PDG} Particle Data Group (D.E. Groom et al.), Eur. Phys. J. C15 
(2000) 1. 
%
\bibitem{lykken} H. Baer, J. Sender and X. Tata, Phys. Rev. D50 (1994) 4517;
R. Demina, J. Lykken, K. Matchev and A. Nomerotski, Phys.\, Rev.\, D62 (2000) 
035011.
%
\bibitem{prod1} G. Kane and J.P. Leveille, Phys. Lett. B112 (1982) 227;
P.R. Harrison and C.H. Llewellyn-Smith, Nucl.Phys. B213 (1983) 223;
C. Reya and D.P. Roy, Phys. Rev. D32 (1985) 645;
S. Dawson, E. Eichtein and C. Quigg, Phys.~Rev. D31 (1985) 1581;
H. Baer and X. Tata, Phys. Lett. B160 (1985) 159. 
%
\bibitem{prod2} W. Beenakker, M. Kramer, T. Plehn, M. Spira and P.M. Zerwas,
Nucl. Phys. B515 (1998) 3; W. Beenakker et al.,  hep-ph/9810290.  
%
\bibitem{CTEQ} CTEQ Collaboration (H.L. Lai et al.), Phys. Rev.D55 (1997) 1280. 
%
\bibitem{ttbar} E. Berger and H. Contopanagos, Phys. Rev. D54 (1996) 3085;
S. Catani, M. Mangano, P. Nason and L. Trentadue, Phys. Lett. B378 (1996) 329.
%
\bibitem{Hagiwara} B.K. Bullock, K. Hagiwara and A.D. Martin, Nucl. Phys. 
B395 (1993) 499. 
%
\bibitem{fake} See for instance, J. Lykken and K. Matchev, Phys.\, Rev.\, 
D61 (2000) 015001  and references therein. 
%
\bibitem{private} G. Bernardi and A. Savoy-Navarro, private communications.
%
\end{thebibliography}
\end{document}